\documentclass[a4paper,10pt]{article}
\usepackage{latexsym, amsmath, amsfonts, amssymb, array, dcolumn, hhline, longtable, enumerate,psfrag,bm}
\usepackage[english]{babel}
\usepackage[dvips]{graphicx}
\usepackage{epsfig}
\usepackage[small,bf]{caption}
\begin{document}
%
%
%

\noindent
{\bf \Large{Solvent Driven Formation of
Bolaamphiphilic Vesicles}}

\vspace{0.5 cm} 

\noindent
{\bf \large {M. I. Katsnelson and A. Fasolino$^*$}}

\vspace{0.5 cm}

\noindent
{\it Institute for Molecules and Materials,
Radboud University Nijmegen,
Toernooiveld 1,
6525 ED Nijmegen, ~~~~~~~~~ ~~~~~~~~~ ~~~~~The Netherlands}

\begin{abstract}
We show that a spontaneous
bending of single layer bolaamphiphiles results from  
the frustration due to the competition
between core-core and tail-solvent interactions. We find that spherical 
vesicles are stable under rather general
assumptions on these interactions described within the Flory-Huggins theory.
We consider also the deformation of the vesicles in 
an external magnetic field that has been recently experimentally observed. 
\end{abstract}


Molecular aggregates, like biological matter, possess the ability
to organize into well defined mesoscopic structures, such as
layers, fibers, vesicles\cite{Jones,Nelson,Elemans,Science}.
Addressing this issue for molecular aggregates is relevant not
only for applications in pharmacology,
catalysis, and other fields, but also to understand the behaviour
of model systems simpler but related to biological matter.
The self-organized
structures result from the concomitant effect of non-covalent
interactions, such as hydrogen bonding and $\pi-\pi$ stacking
between molecules, as well as by entropic contributions. A theory
capable of predicting the shape of the aggregate from the
knowledge of the microscopic molecular structure is still missing.
Among the forms of self-organization, the formation of empty
vesicles in, possibly aqueous, solution is particularly important 
for the wealth of possible applications, from
microreactors to drug delivery. Surprisingly, vesicle formation has been
observed not only in the well known case of amphiphiles bilayers,
like lipid membranes\cite{Jones,Nelson,Seifert} but also in several
bolaamphiphiles\cite{Matsuzawa,Yan,Igor}
formed by an hydrophobic core and two, usually symmetric,
lateral hydrocarbon tails terminated by hydrophilic
groups\cite{Fuhrhop}. Notice that the term bolaamphiphiles has been first introduced to describe synthetic analogs of archaebacterial membranes\cite{Fuhrop2,Luzzati}.
Here we provide a model predicting spontaneous bending due to frustration
resulting from competing core-core and tail-solvent interactions. 

\begin{figure}[h]
 \begin{center}
   \centering
   \includegraphics[width=14.cm,keepaspectratio]{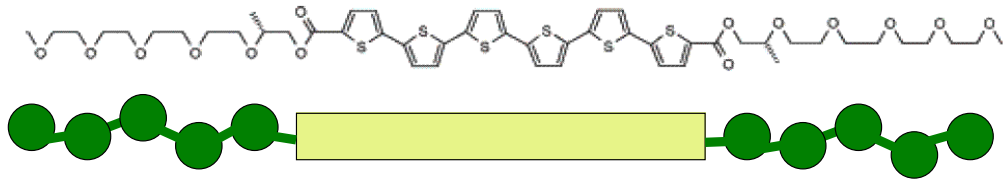}
   \caption{Structural formula of 2,5'''''-(R-2-methyl-
3,6,9,12,15-pentaoxahexadecyl ester) sexithiophene and its schematic representation. }
            \label{figsexi}
 \end{center}
\end{figure}

We refer, in particular, to the case of 
sexithiophenes bolaamphiphiles (see Fig.~\ref{figsexi}) that have been found to form vesicles
albeit
in isopropanol\cite{Igor}. The formation of stable vesicles seems to be
favoured by increasing length of the tails\cite{Yan}.
Conversely, layer-like structures have been found
in liquid crystal bolaamphiphiles with short or rigid lateral
blocks\cite{Cheng2002,Cheng2004}.
\vspace{1.cm}

\begin{figure}[h]
 \begin{center}
   \centering
   \includegraphics[width=8.cm,keepaspectratio]{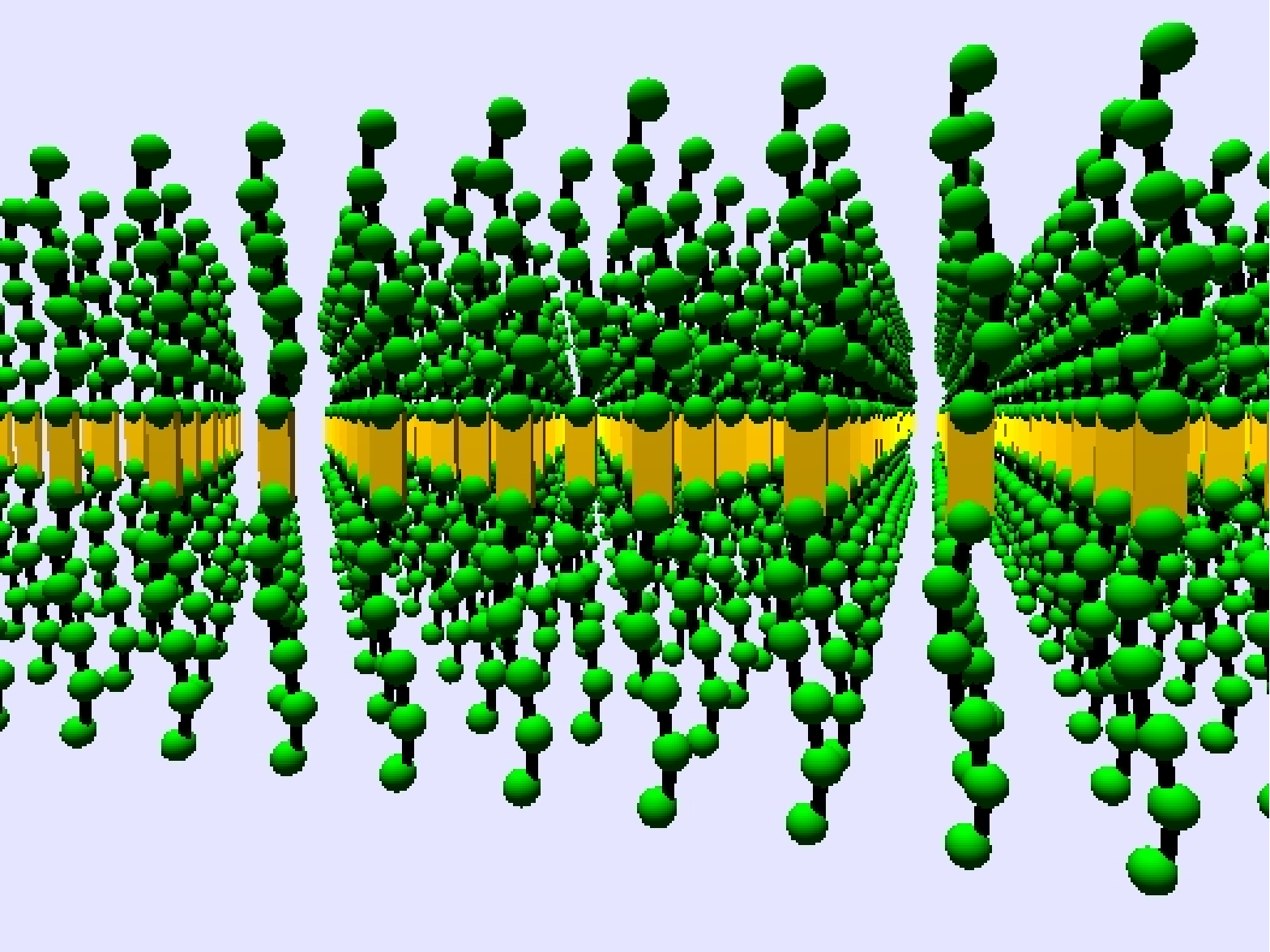}
   \caption{Sketch of a layer formed by self-assembled bolaamphiphiles.
The spacing between hydrophilic tails (green) is fixed by the hydrophobic cores (yellow) to a value which might be not
optimal for tail-solvent interactions.}
            \label{figure_1}
 \end{center}
\end{figure}

We consider a simplified model of a bolaamphiphile layer with the cores
on a two-dimensional surface and symmetric flexible tails, each
formed by $m$ monomers as sketched in Fig.\ref{figure_1}. 

\begin{figure}[h]
 \begin{center}
   \centering
   \includegraphics[width=8cm,clip]{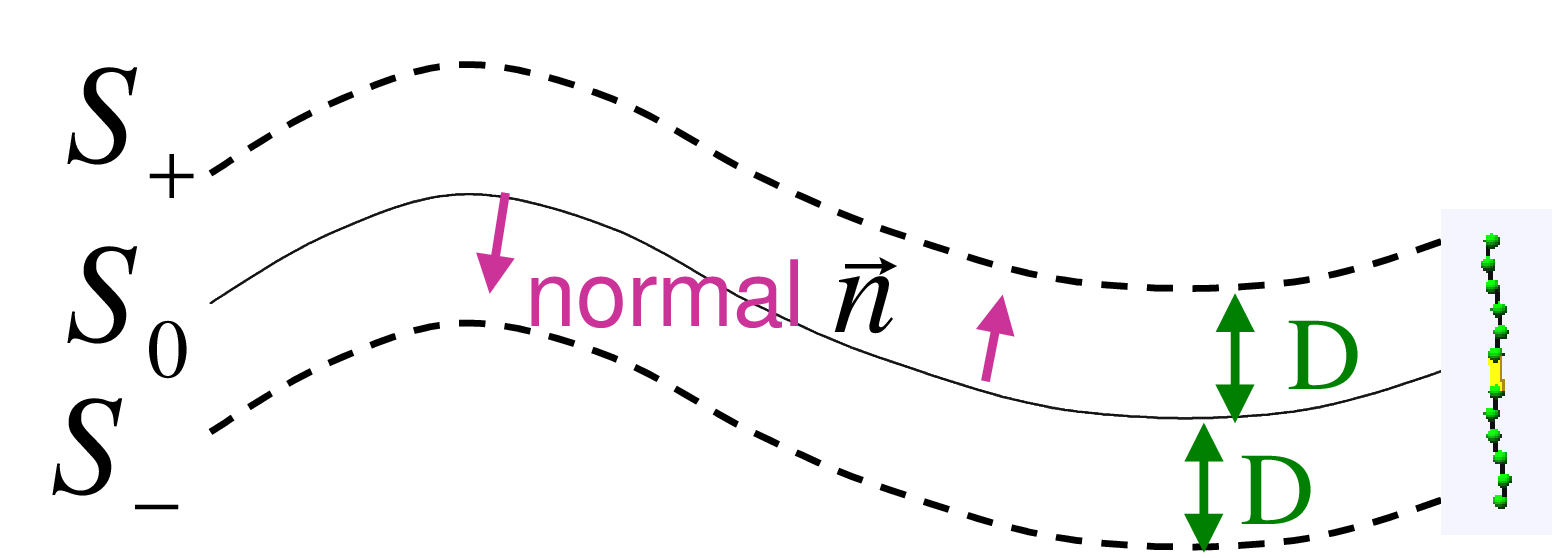}
   \caption{Schematic geometry of our model. The surface $S_0$ is the locus of the center of mass of the hydrophobic cores and $S_\pm$ are the surfaces formed by the tail ends. The unit vector $\vec{n}$ is normal to all three surfaces and  $D$ is the effective tail length. }
            \label{figure_2}
 \end{center}
\end{figure}

As illustrated in Fig.\ref{figure_2}, the cores define a surface $S_0$
with local coordinates $(u_1,u_2)$ such that
$\vec{r}_0=\vec{r}(u_1,u_2)$ gives cartesian coordinates on the
surface. 
If the tails, due to covalent bonding to the cores, tend
to be perpendicular to the surface, the interactions with the
solvent are supposed to take place on the two surfaces $S_{\pm}$
with coordinates
\begin{equation}
\vec{r}_{\pm}=\vec{r}(u_1,u_2)\pm \vec{n}(u_1,u_2)D
\end{equation}
where $\vec{n}$ is the vector normal to $S_0$ and $D$ is the
effective length of the tails.

The elementary area vector on $S_0$ is
\begin{equation}
d\vec{S}_0=\left( \vec{r}_1\times \vec{r}_2 \right) du_1 du_2
\end{equation}
where $\vec{r}_i=\partial{\vec{r}}/\partial{u_i}$. The derivative
of the normal $\vec{n}$ with respect to $u_i$
is given by the so-called Weingarten
equations\cite{Coxeter}, whence
\begin{equation}
d\vec{S}_\pm=d\vec{S}_0\left( 1 \mp 2HD + KD^2 \right)
\end{equation}
where $H=(\kappa_1+\kappa_2)/2$ and $K=\kappa_1\kappa_2$ are,
respectively, the mean and gaussian curvature defined in terms of
the principal curvatures $\kappa_1$ and $\kappa_2$. 
The tail density in contact with the solvent is
determined by the tail density $n_0$ imposed by the core so that $n_\pm d\vec{S}_\pm= n_0
d\vec{S}_0$ leading to an exact expression for the tail
density on $S_\pm$ in terms of the curvature 
\begin{equation}
n_\pm= \frac{ n_0}{ 1 \mp 2HD+KD^2}.
\end{equation}
If the tails were not rigidly connected to the cores, they would
form a solution with free energy per unit volume $\phi(n)$.
Therefore, we write the bending free energy  as
\begin{equation}
\delta F=D \int dS \left[\phi(n_+)+\phi(n_-)\right]+ A \int dS H^2
+ B \int dS K. \label{deltaF}
\end{equation}
The first term is the free energy of the tail/solvent subsystem
whereas the terms proportional to $A$ and $B$ represent  
the bending elasticity of the core
layers\cite{Jones,Nelson,Seifert}. 
To account phenomenologically
for bent surfaces for lipid bilayers, the term $ A H^2$ is usually
replaced by the so-called Canham-Helfrich (CH) bending energy $ A
(H-H_0)^2$ with $A>0$ so that negative linear terms in $H$ are
responsible for bending\cite{Jones}.  However, the CH expression is
not appropriate for symmetric bolaamphiphiles where odd terms
vanish by symmetry. Moreover, the spontaneous curvature $H_0$
is introduced phenomenologically and is not justified by any
microscopic model.  We also note that the term
$B \int dS K$ is usually neglected because due to the Gauss-Bonnet
theorem\cite{Coxeter} it is constant within a given
topology of the surface and it actually depends on the number of
handles. 
We consider mesoscopic
vesicles with curvature much larger than the molecular size,
namely $ |\kappa_i|D<<1$. Within this approximation we can expand
$\phi(n_\pm)$ up to second order in $D$ as
$$
\phi(n_\pm)=\phi(n_0)\pm n_0\phi'(n_0)2HD+ ~~~~~~~~~~~~~~~~~~~~~~~~~
$$
\begin{equation}
+ n_0\phi'(n_0)\left[4H^2D^2-KD^2\right]
+\frac{1}{2}n_0^2 \phi''(n_0)4H^2D^2
\end{equation}
where $\phi'(n_0)=d\phi/dn|_{n=n_0}$ and $\phi''(n_0)=d^2\phi/dn^2|_{n=n_0}$.
Notice that the sum of the terms linear in $H$ in the integrand of 
Eq.\ref{deltaF} vanishes
whereas other terms contribute to the term proportional
to $H^2$  so that
\begin{equation}
A \rightarrow \tilde{A}=A+\delta A=A +4D^3\left[ n_0^2\phi''+2 n_0 \phi'\right]
\end{equation}
We will show that $\tilde{A}$ can become negative and lead to
spontaneous bending as a result of the microscopic tail/solvent
interactions. 
The reason for a negative
$\delta A$ is that the available volume per tail on the two sides
of a bent surface may be different and a specific curvature can
minimize the resulting free energy.

We can further specify the driving term $\delta A$ by calculating
it, within the Flory-Huggins theory for polymers solvent
mixtures\cite{Flory,Jones}
$$
\frac{\phi(n)}{T}= \chi \frac {m n n_1}{n_1+m n} +
$$
\begin{equation}
+ n_1 \ln{\left(\frac{n_1}{n_1+mn}\right)}
+ n_2 \ln{\left(\frac{mn}{n_1+mn}\right)}
\end{equation}
where $n_1$ is the solvent density, $m$ is the number of monomers per tail,
$\chi\sim (2W_{ts} -W_{ss}-W_{tt})/T$ with $W_{ts}$, $W_{ss}$ and
$W_{tt}$ the tail-solvent, solvent-solvent and tail-tail
interaction energies respectively and $T$ is the temperature.
Substitution of Eq.8 into Eq.7 gives
$$
\frac{\delta A}{T} = \frac{2 \chi m n_1^3 n_0}{ \left( n_1+mn_0 \right)^2} -
\frac{n_1}{\left(n_1+m n_0\right)^2} \cdot 
$$
\begin{equation}
\cdot \left[m \left(m-2\right)n_0+\left(2m-3\right)n_1\right]
\end{equation}
For hydrophilic tails, 
$\chi<0$
leading to a free energy gain for bent surfaces. This constitutes a 
possible microscopic mechanism for vesicle formation, albeit within the 
Flory-Huggins theory which is more qualitative than quantitative\cite{Koningsveld}.
Furthermore, the last term that accurately represents entropy is
always negative unless $m=1$ and increases as $m^2$ for $m \gg 1$. The
observation of planar structures in liquid crystals
bolaamphiphiles with rigid tails could be explained by this
entropic reason. 
This completes our microscopic analysis. Our
model leads naturally to a negative contribution to the free
energy of the term quadratic in the curvature that does not need
to be imposed phenomenologically as  done for amphiphile bilayers.

Our results can be cast in the form of an effective Landau
Hamiltonian, i.e. we can write the free energy $F$ as a Taylor expansion 
in terms of the curvatures $H$ and $K$. These quantities, or more precisely $\kappa_1$ and $\kappa_2$, take the apparent role of order parameters assuming their smallness in comparison to a typical inverse molecular size. However, $\kappa_1$ and $\kappa_2$ are not
independent at different points of the surface, implying that
$F$ is not a true Landau Hamiltonian and that a local analysis
is insufficient. In fact, the local free energy density has not
a minimum but a saddle point for a sphere ($\kappa_1=\kappa_2$)
if $c_2+2c_3<0$, in contradiction with the
numerical data presented below. Since the term proportional to
$H^2$ resulting from our microscopic model is negative, higher
order terms are needed to stabilize the mean curvature to a finite
value. Up to fourth order terms in $\kappa_1$ and $\kappa_2$ we
can write
\begin{equation}
F=\int {dS \left[ -|A|H^2 +bHK+b'H^3+c_1H^4+c_2H^2K+c_3K^2\right]}
\label{F(H,K)}
\end{equation}

The odd terms in $b$ and $b'$ that can lead to first order phase
transitions have to vanish for the symmetric bolaamphiphiles
considered here so that we take $b=0$ and $b'=0$. Moreover,
$c_1+c_2+c_3>0$. Since, in
general, the bending energy is much smaller that the surface
tension, we go to the limit of infinite surface tension that
allows to perform minimization of  $F$ on a surface of constant
area. We have minimized Eq.\ref{F(H,K)} numerically for surfaces
topologically equivalent to spheres with radius
$R=\sqrt{2\left(c_1+c_2+c_3\right)/|A|}$ and found that spheres 
are always the solution.

\begin{figure}[h]
 \begin{center}
   \centering
   \includegraphics[height=8.cm,keepaspectratio,angle=-90]{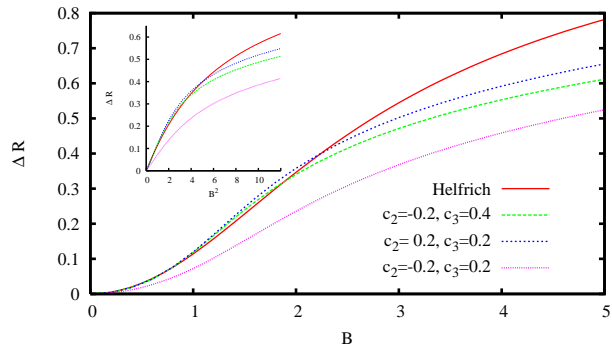}
   \caption{Difference $\Delta R$ between short and long axis of the ellipsoid resulting from minimization of the free energy in a magnetic field $B$ for three choices of $c_2$ and $c_3$ ($c_1+c_2+c_3=1$ and 
$|A|=1$), compared with the Helfrich model ($H_0=1$). The inset shows that initial $B^2$ dependence of the distortion depends mostly on 
$c_2+2c_3$.}
            \label{figure_3}
 \end{center}
\end{figure}

Our model can be extended to consider the recently observed effect
of external magnetic fields $B$ on the equilibrium shape of
bolaamphiphilic vesicles\cite{Igor}. To this purpose we  add
to Eq.\ref{F(H,K)} the diamagnetic energy term
\begin{equation}
E_{mag}=-\sum_i\frac {D_i B^2}{2\mu_0}\int {dS \left( \chi^i_\perp \sin^2\theta
+\chi^i_\parallel \cos^2\theta \right)}
\label{Emag}
\end{equation}
where $D_i$ is the effective thickness of the core layer ($i=1$) or 
of the tails ($i=2$), 
$\chi^i_\perp$, $\chi^i_\parallel$ are
the components of the corresponding magnetic susceptibility  and $\theta$ is the polar angle of the direction of the magnetic field $B$.
Due to the tendency of the aromatic rings in the core segment to align 
parallel to the magnetic field, the spherical vesicles
will be transformed into
ellipsoids, as predicted by Helfrich\cite{Helf} for amphiphilic bilayers. 
This deformation in high magnetic fields has been recently observed experimentally 
for sexithiophene\cite{Igor}, that is an example of symmetric 
bolaamphiphiles\cite{Igor} where the CH model of free energy is not correct. 
In Fig.\ref{figure_3} we present the deformation calculated by minimizing the sum of Eqs. 10 and 11
for constant surface as a function of magnetic field, compared to the one resulting 
from the CH model. In both cases,  
the deformation is proportional to $B^2$ for low fields and flattens out as the field increases. However, for the same initial slope, our model predicts smaller deformations at high fields, the details of the curve being determined by 
the parameters $c_1, c_2, c_3$.  

In summary, we have demonstrated that the frustration resulting from 
competing core-core and tail-solvent interactions can lead to
spontaneous bending of  single bolaamphiphilic layers. By describing the
tail solvent interactions within the Flory-Huggins theory, 
we have constructed a Landau-like free energy appropriate to describe 
symmetric bolaamphiphiles that gives a rationale for the formation of spherical vesicles. Measurement of their deformation in high magnetic
fields can provide information about the parameters of the theory, opening the possibility to validate microscopic models of interactions in these systems. 


{\bf Acknowledgment} ~~~We thank Igor Shklyarevsky, Peter Christianen, Jan Kees Maan, Ad van der Avoird and Roeland Nolte for useful discussions.

\vfill\eject
\newpage
\begin{figure}[t]
\begin{minipage}[c]{0.4\linewidth}
   \includegraphics[width=5cm,keepaspectratio]{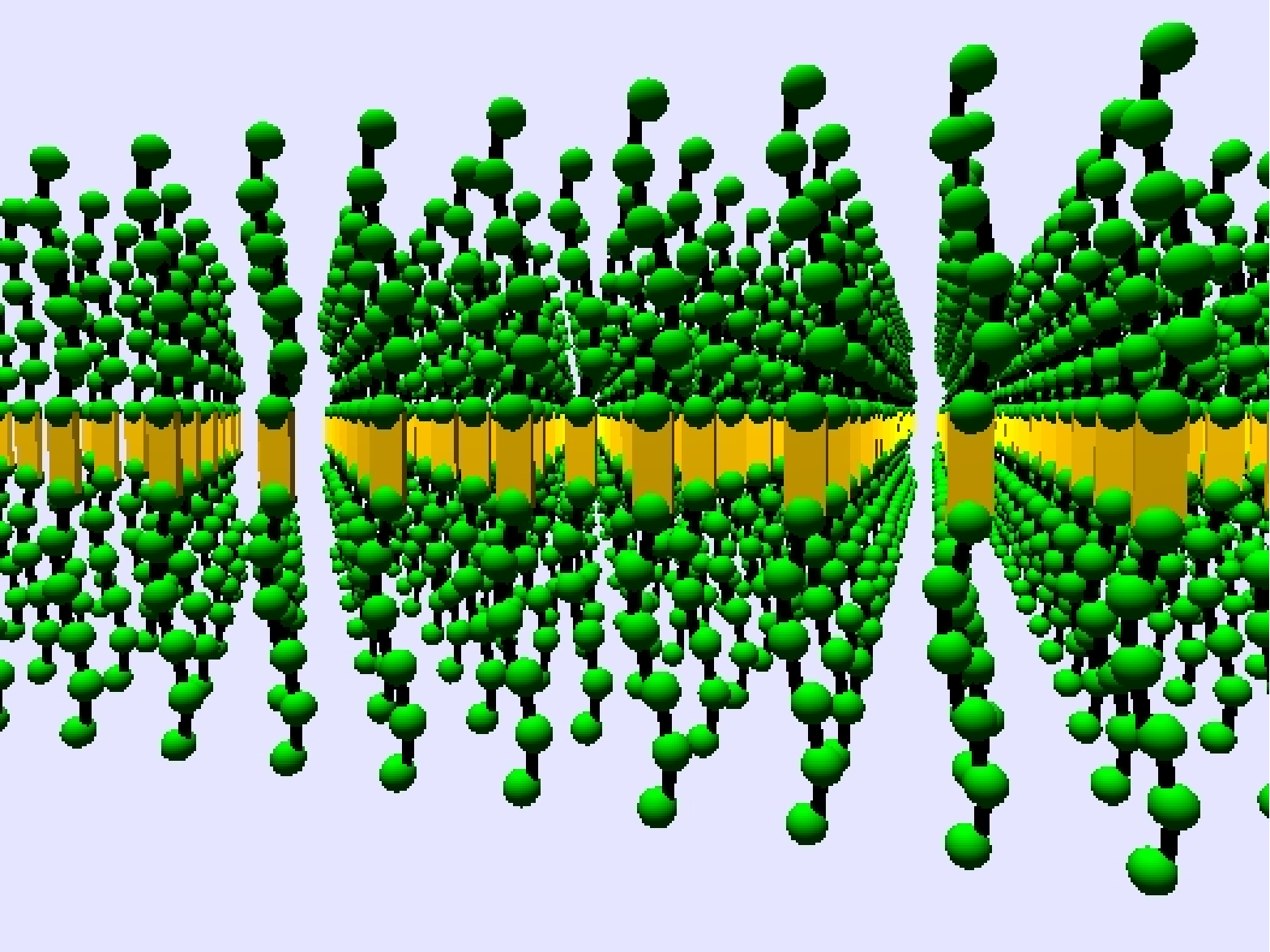}
\end{minipage}
\hskip -0.2cm
\begin{minipage}[c]{0.55\linewidth}
   \includegraphics[width=7cm,keepaspectratio]{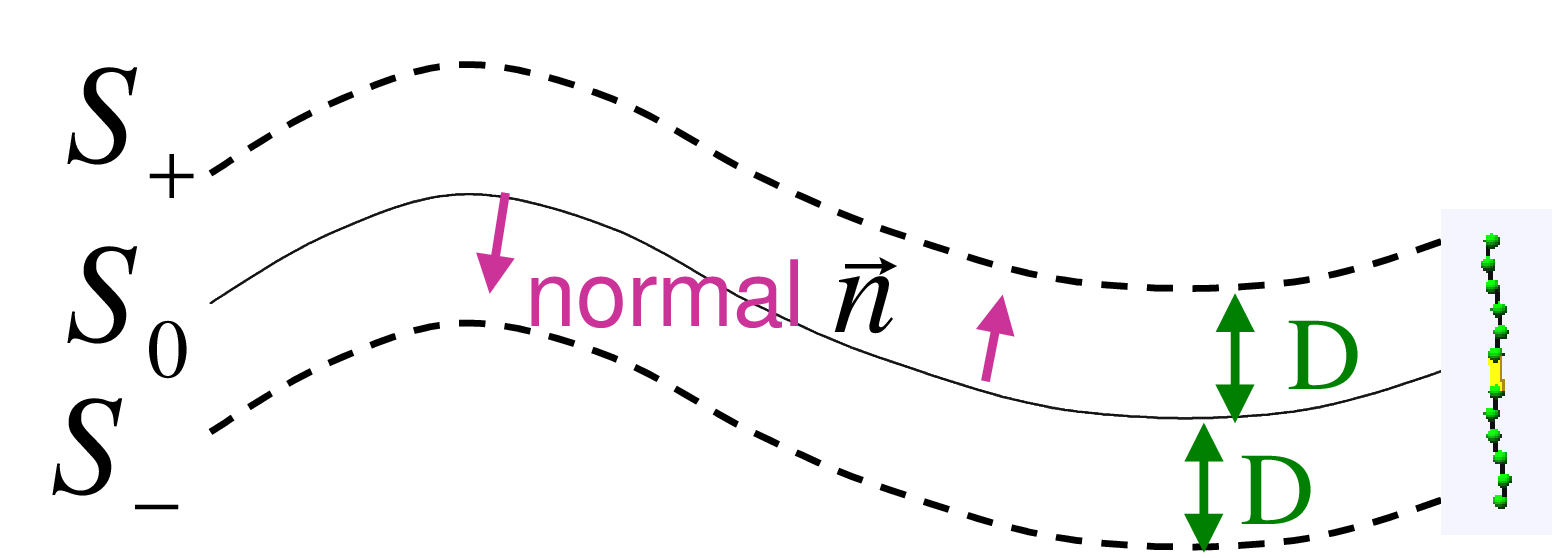}
\end{minipage}

\end{figure}

\end{document}